# *Composites of FeCl$_3$ and TiO$_2$ with Bismaleimide resins*

**A.S.Bhattacharyya**[1, 2*], **Nitesh Rajak**[1], **Abhishek Sharma**[1], **P. Kommu**[1]

[1]Centre for Nanotechnology
[2] Centre of Excellence in Green and Efficient Energy Technology (CoE-GEET)
Central University of Jharkhand, Ranchi – 835 205, India,
E-mail: 2006asb@gmail.com ; arnab.bhattacharya@cuj.ac.in

**Abstract**

Ferric Chloride-Bismaleimide (FeCl$_3$-BMI) and Titania-Bismaleimide (TiO$_2$-BMI) composite were synthesized using PVA as a binder. The composite systems were deposited on glass slide as a homogenous coating. XRD and FTIR studies of the composite system showed its crystalline and structural configuration. A mixed phase of TiO2 and BMI as well as short range crystallinity was observed. An increase in crystallinity with temperature was also seen. The percentage of N-H symmetric stretching was also found to increase with temperature.

## Introduction

Bismaleimide are thermosetting resins base on aromatic diamines. Addition of inorganic materials like metal oxides can influence the properties of the polymer as an inorganic-organic composite. Bismaleimide (BMI) on the other hand are thermosetting polymers having properties of dimensional stability, low shrinkage, chemical resistance, fire resistance, good mechanical properties and high resistance against various solvents, acids, and water [1, 2]. BMI is commercially available in different forms [3]. BMI coating has also been used for the corrosion protection [4]. Metallization as well as surface properties of BMI are published elsewhere [5, 6]. It has applications in airforce, military and electronics [7,8]. Addition of inorganic materials like metal oxides like TiO$_2$ can influence the properties of the polymer as an inorganic-organic composite. FeCl$_3$ on the other hand has been used previously in making Graphite Nanosheet (GNS)-Fe$_3$O$_4$-BMI composites [9].

## BMI Compounds

The basic structure of BMI is given in Fig 1 where the R can vary to form other different compounds of BMI like sulphone and sulphone-ether BMI, ether and ether ketone BMI as given by Wilson et. al [10].

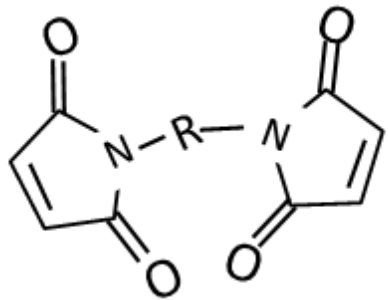

Fig 1: Basic structure of BMI [10] **(self drawn)**

BMI bas is available in different forms viz : MDA-BMI (methylene dianiline bismaleimide), MPD-BMI (m-phenylene diamine bismaleimide), PPD-BMI (p-phenylene diamine bismaleimide), TDA-BMI (toluene diamine bismaleimide) BMI-70 (3,3'-diethyl-5,5'-dimethyl methylene dianiline bismaleimide) and BMI-80 (2,2-bis(4[4-aminophenoxyl]phenyl)propane bismaleimide) according to Rojstaczer et. al [11].

According to Donnelan et al, although BMI are brittle on curing which leads to impact damage in composite laminates, a slower curing rate causes enhanced chain extension (called *Michael addition*) prior to cross linking which may improve the fracture toughness. The cross linking density was found to be the most important parameter while studying thermal and mechanical properties. Fracture toughness increased with decrease in cross linking density [12, 13].

## Experimental

$TiO_2$-BMI and $FeCl_3$-BMI composite systems were made by taking $TiO_2$ and $FeCl_3$ powder, PVA and BMI (10 gm) and using magnetic stirrer. The system was quite homogenous and had no precipitation. A coating of the composite was made by uniformly depositing on a glass slide and allowing it to dry (Fig 1). Proto XRD at Centre of Excellence for Green Energy and Efficient Technology (CoE-GEET) CUJ was used for the XRD studies where monochromatic Cu-K$_\alpha$ radiation having a wavelength of 1.54Å was used.

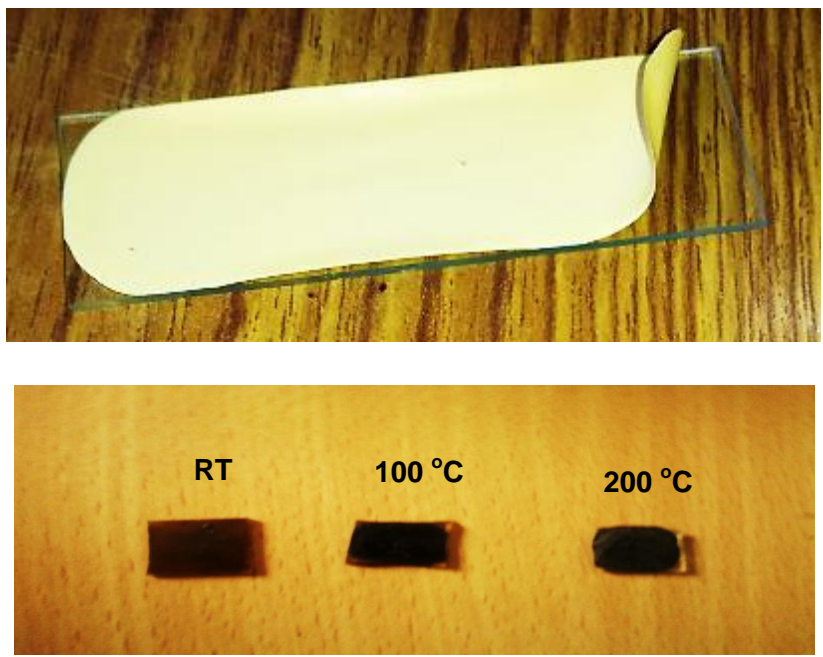

**Fig 1**: a) $TiO_2$-BMI and b) $FeCl_3$-BMI composite on glass slide

## Results and discussions

XRD of $FeCl_3$-BMI composite systems synthesized at different temperatures are shown in fig 1. Increase in temperature led to an increase in crystallinity. A reaction of the amine group with the malemide double bond followed by cross-linking takes place takes place with increase in temperature [12].According to Chandran M et al alignment of segments along their axis during chain-chain interactions during intermolecular charge transfer process causes crystallinity in BMI resins [14]. However the use of FeCl3 seems to aid in the crosslinking and charge transfer to a great extent and alignment only in a particular direction giving rise to very sharp peak which also found to increase with temperature.

FTIR spectra of the system synthesized at $100^oC$ and $200^oC$ was done as shown in Fig 2. A broad peak around 3500 $cm^{-1}$ and sharp peak at 1625 $cm^{-1}$ were observed due to N-H stretching and bending respectively [15]. The peak around 3500 $cm^{-1}$ was deconvoluted as shown in Fig 3 and 4 with parameters given in Table 1. A sharper and more intense peak was observed with increase in temperature. Primary amines have two bands between 3500-3300 $cm^{-1}$ due to asymmetric and symmetric stretching and a band between 1640 -1560 $cm^{-1}$ due to bending [16]. The asymmetric stretching usually takes place at lower wave number but of higher intensity and magnitude compared to symmetry stretching. The ratio of intensity, area and width of the asymmetric and symmetric modes were calculated. It can be observed that with increase in temperature, the intensity of symmetric N-H stretching increases.

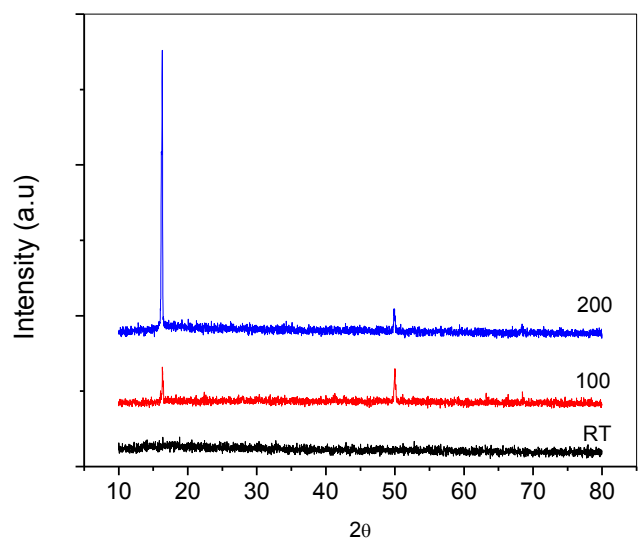

**Fig 1**: XRD of FeCl$_3$-BMI composite synthesizes at different temperatures

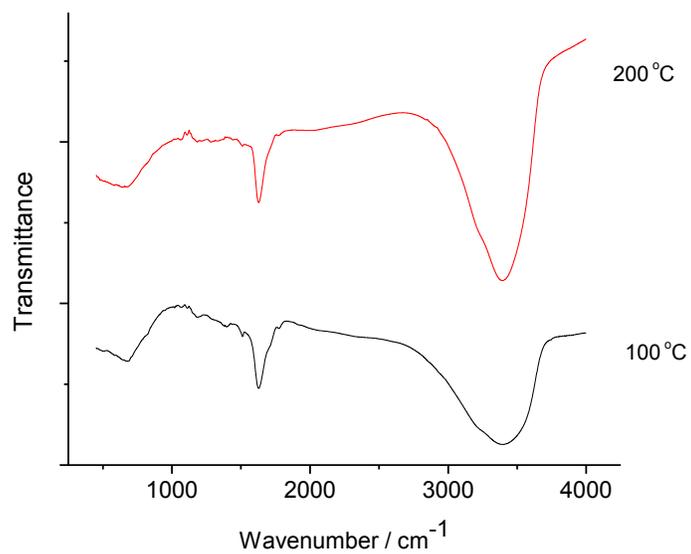

**Fig 2**: FTIR of BMI-FeCl$_3$ synthesized at 100 and 200 °C

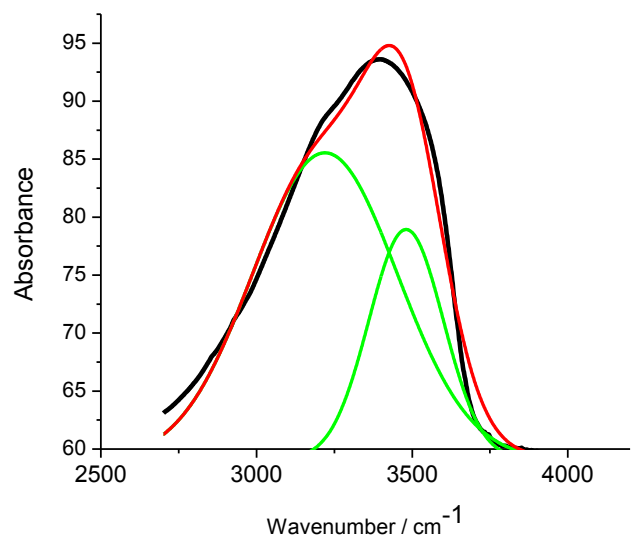

**Fig 3**: Deconvoluted peaks of BMI-FeCl$_3$ synthesized at 100 $^o$C

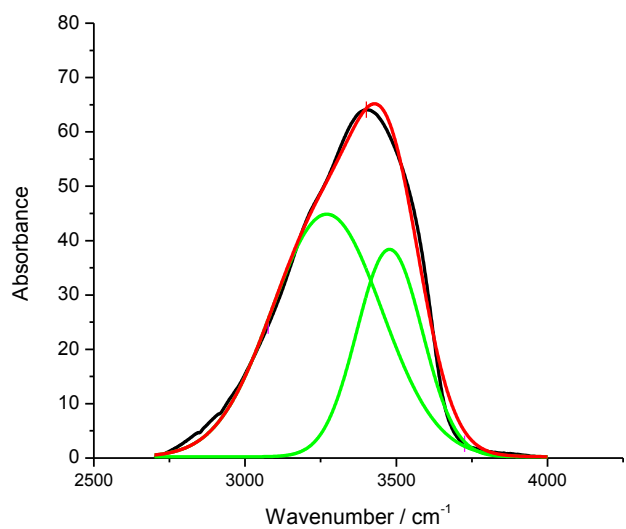

**Fig 4**: Deconvoluted peaks of BMI-FeCl$_3$ synthesized at 200 $^o$C

**Table 1**: Fitting parameters of deconvolution of the FTIR spectra

| Sample | Peak | N-H stretching | Height | Height ratio | FWHM | FWHM ratio | Area | Area ratio |
|---|---|---|---|---|---|---|---|---|
| BMI-F 100 °C | 3220 | asymmetric | 26.5 | 1.34 | 464 | 1.94 | 15395 | 2.59 |
|  | 3481 | symmetric | 19.8 |  | 239 |  | 5941 |  |
| BMI-F 200 °C | 3270 | asymmetric | 45 | 1.18 | 364 | 1.68 | 20330 | 1.96 |
|  | 3478 | symmetric | 38 |  | 216 |  | 10344 |  |

XRD of $TiO_2$, BMI and $TiO_2$-BMI is shown in Fig 5. XRD of the mixed phase showed contribution from both the phases. The XRD of PVA usually shows a peak at 20°. However there is also a major contribution of BMI at the same angle. Since PVA was taken in very small quantity, we assume that the broad peak at 20° is due to BMI only. The sharp diffraction peaks of the unmodified BMI between 10° to 30° indicate that it's a brittle resin [14]. Electronic polarization between diamine and dianhydride regions and consequently a phenomenon called charge transfer complexion occurs in BMI. The alignment of segmented rigid rod polymides due to chain-chain interactions forms short range order and crystallinity as in the previous case. The sharp peaks in the region 10° to 30° disappear in the modified ($TiO_2$-BMI) system indicating loss in crystallinity. The effect of temperature and APPJ treatment on crystallinity has been published elsewhere [15, 17]. Interestingly, the nature of crystallinity in $TiO_2$-BMI and $FeCl_3$-BMI was completely different which requires further investigations.

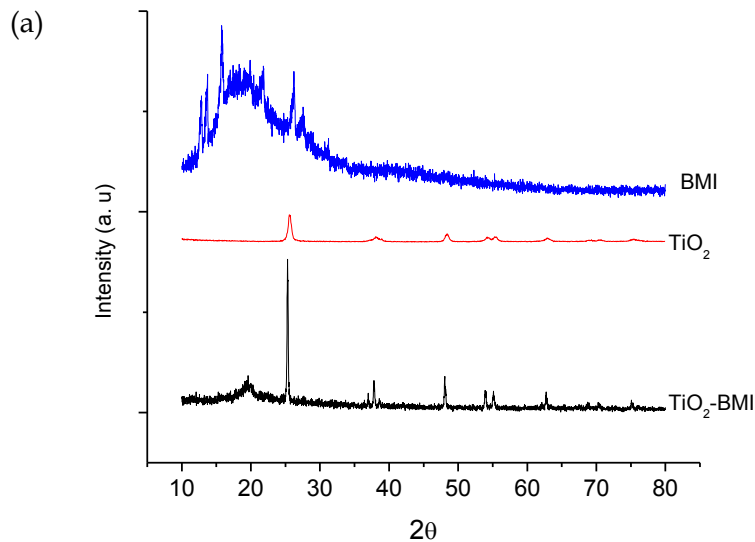

(a)

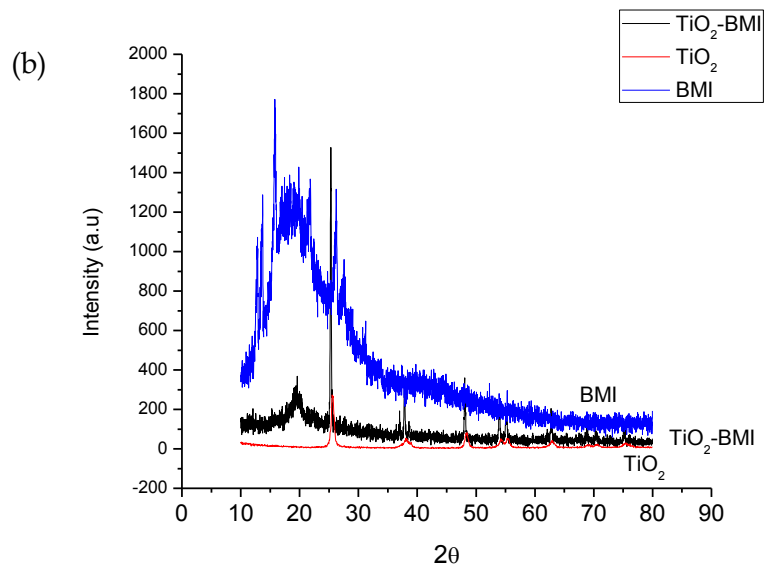

**Fig 5 :** XRD of TiO$_2$, BMI and TiO2-BMI   (a) Y offset (b) normal mode

## Conclusions

Ferric Chloride-Bismaleimide (FeCl$_3$-BMI) Titania-Bismaleimide (TiO$_2$-BMI) and composites synthesized showed increased crystallinity with increase in temperature. FeCl$_3$ seems to aid in the charge transfer and alignment process to a great extent. XRD showed the mixed phases of TiO$_2$ and BMI as well as short range order and crystallinity FTIR studies showed increase in N-H symmetric stretch concentration compared asymmetric stretch with increase in temperature.

## Acknowledgements

The authors thank Science and Engineering Research Board, India for research grant SERB/F/3482/2012-2013 (Dated 24 September 2012) and Centre of Excellence in Green and Efficient Energy technology (CoE-GEET) under FAST Scheme, MHRD India**.** The authors would also like to thank Pallavi of Centre for Applied Chemistry, Central University of Jharkhand for FTIR studies.